\documentclass[twocolumn,prl,amsmath,amssymb,showpacs,superscriptaddress]{revtex4-1}
\usepackage{epsf}      
\usepackage{graphicx}
\usepackage{color}
\usepackage{soul}
\usepackage{gensymb}

\begin{document}

\title {Physical properties and electrochemical performance of Zn-substituted Na$_{0.44}$Mn$_{1-x}$Zn$_x$O$_2$ nanostructures as cathode in Na-ion batteries}

\author{Mahesh Chandra}
\affiliation{Department of Physics, Indian Institute of Technology Delhi, Hauz Khas, New Delhi-110016, India}
\author{Rishabh Shukla}
\affiliation{Department of Physics, Indian Institute of Technology Delhi, Hauz Khas, New Delhi-110016, India}
\author{Rakesh Saroha}
\affiliation{Department of Physics, Indian Institute of Technology Delhi, Hauz Khas, New Delhi-110016, India}
\author{A. K. Panwar}
\affiliation{Department of Applied Physics, Delhi Technological University, Delhi-110042, India}
\author{Amit Gupta}
\affiliation{Department of Mechanical Engineering, Indian Institute of Technology Delhi, Hauz Khas, New Delhi-110016, India}
\author{S. Basu}
\affiliation{Department of Chemical Engineering, Indian Institute of Technology Delhi, Hauz Khas, New Delhi-110016, India}
\author{R. S. Dhaka}
\email{rsdhaka@physics.iitd.ac.in}
\affiliation{Department of Physics, Indian Institute of Technology Delhi, Hauz Khas, New Delhi-110016, India}

\date{\today}                                         

\begin{abstract}
We report the synthesis, physical properties and electrochemical performance of Zn substituted Na$_{0.44}$Mn$_{1-x}$Zn$_x$O$_2$ ($x=$ 0 -- 0.02) nanostructures as cathode in Na-ion batteries for energy storage applications. These samples stabilize in the orthorhombic structure and the morphology is found to be slab like with 100 -- 200~nm width and few micrometer of length. The resistivity measurements show highly insulating nature for all the samples, where the activation energy decreases with increasing Zn concentration indicating more defect levels in the band gap. The cyclic voltammogram (CV) shows reversible oxidation and reduction peaks, which clearly shift towards higher/lower potentials with increasing Zn concentration up to 2\%. We observed the specific capacity of about 100~mAh/g at current density of 4 mA/g and improved cycle life for Zn substituted $x=$ 0.005 sample. However, with further increasing the Zn concentration ($x=$ 0.02), the specific capacity decreases, which can be a manifestation of the decreased number of reduction peaks in the CV data. 
\end{abstract}

\maketitle

\section{\noindent ~Introduction}

 All electrochemical intercalation based batteries require not only ample lattice space and robust structure of an electrode, but also a multiple valance ion to maintain charge neutrality \cite{Rev1}. In this regard transition metal based oxides are widely studied as cathode as well as anode in Li/Na/K ion batteries \cite{Rev1, JameshJPS18, Rev TMO, revNMO2017, XiaoAEM18}. Among transition metal oxides, Na$_{0.44}$MnO$_2$ with tunnel type structure and multivalent Mn ion have been studied as cathode material (in pristine form) \cite{NM44synRev, ChenACSAMI18} as well as anode material (when doped with other appropriate transition metal ion) in a Na-ion battery \cite{Ti NMO}. The 3D lattice structure of Na$_{0.44}$MnO$_2$ comprise of MnO$_6$ octahedra and MnO$_5$ square based pyramids with 0.44 Mn in 3+ valence state and  0.56 Mn in 4+ valence state. Among these, all the Mn$^{4+}$ and 50\% Mn$^{3+}$ occupy the octahedral position whereas rest of the 50\% of Mn$^{3+}$ occupy pyramidal position resulting into three different sites for Na ions in two different tunnels, one large S-shaped and other smaller one \cite{SauvageInChem2007}. These tunnels in Na$_{0.44}$MnO$_2$ allow an easy insertion and reinsertion of the Na--ions during the oxidation and reduction of the cathode material.
 
The electrochemical properties and particularly the cycling capacity of Na$_{0.44}$MnO$_2$ critically depend on the synthesis method \cite{SauvageInChem2007,tempAssSolGel2016,polypyrolysisNanowire}, sintering temperature \cite{SauvageInChem2007,NM44synRev} and morphology \cite{Nanorods2016,DemirelMT,ZhaoRSC13Raman} of the samples. Despite of the potential, two major challenges need to be overcome while using these materials as cathode in Na-ion batteries. First is the poor retention of capacity upon cycling and low capacity at higher current rates. The poor cyclic capacity is owing to the structural instability upon insertion/extraction of the Na--ions \cite{NM44synRev,SauvageInChem2007}. Such deficiency can be resolved by altering the morphology (surface to volume ratio) by optimizing the growth technique. There have been various modified techniques employed for the synthesis of Na$_{0.44}$MnO$_2$ \cite{ZhangEA17, MaEA16, AstaACSAMI17, JuJPC18}, such as reverse microemulsion method \cite{reverseMicroMethod,NM44synRev}, polymer-pyrolysis method \cite{polypyrolysisNanowire}, modified pechini method \cite{modPechini,ZhaoRSC13Raman} and optimized solid state reaction method \cite{DemirelMT}. Apart from the synthesis, introduction of some disorder in the electrode material, especially in the form of doping, has been found to be increasing the capacity and performance of the electrode material \cite{Rev TMO, Rev1,Ti NMO,Cr dope Nat}.

Another major issue with Mn based transition metal oxides is the presence of the strong Jahn-Teller (JT) active Mn$^{3+}$ ion. The JT distortion makes the compound vulnerable to the structural changes during the Na insertion/extraction resulting into the reduction of capacity after cycling. One approach to overcome this limitation is to substitute the transition metal site with a diavalent ion \cite{Zn doping ref, Na content}, which eventually converts some of the Mn$^{3+}$ to Mn$^{4+}$, therefore reducing the number of JT ions. One such example is substitution of Ni at Mn site where the most stable state of Ni (which is Ni$^{2+}$) is utilized \cite{Na content}. Similarly, the most stable state of Zn is 2+ and therefore, can be a potential candidate for the replacement of Mn. Moreover, the ionic size of Zn is larger than that of Mn, so one can expect an increase in the unit cell volume with Zn substitution and hence easy insertion/reinsertion of Na ions upon charging/discharging. The larger ion in the lattice or larger unit cell volume upon substitution has been found to improve electrochemical properties in some oxides \cite{SarohaAppSurSci2017, LeeSciReP2017}.

Therefore, we report the synthesis of Zn-substituted tunnel type Na$_{0.44}$MnO$_2$ by sol-gel technique and investigation of their structural, morphological, transport and electrochemical properties. These samples stabilize in the orthorhombic structure and the activation energy decreases with increasing Zn concentration. We observed the specific capacity of about 100~mAh/g at current density of 4~mA/g. The capacity does not increase with Zn substitution, but we observed significant improvement in the cycling life of a Na-ion battery.

\section{\noindent ~Experimental}

We have synthesized Na$_{0.44}$Mn$_{1-x}$Zn$_x$O$_2$ ($x=$ 0, 0.005, 0.02) using sol gel method. The sodium nitrate  (Merck, 99\%), manganese nitrate  (Merck, 99\%), and zinc nitrate (Merck, 99.9\%) were added in a stoichiometric ratio in deionized water followed by homogeneouse mixing via stirring for 2~hrs. Then, citric acid (Sigma, 99.9\%) in molar ratio of 3:1 was added to the mixture as a complexing agent followed by overnight stirring at 90$^o$C, which results in the formation of the gel. The formed gel was dried at 120$^o$C for 24~hrs. The obtained powder then ground to get fine particles. The powder was first heated to 450$^o$C for 12~hrs to remove organic components before final sintering at 900$^o$C for 15~hrs. 

The room temperature powder x-ray diffraction data were recorded with CuK$\alpha$ radiation (1.5406~\AA ) from Panalytical x-ray diffractometer in the 2$\theta$ range of 10 -- 60$^o$. The surface morphology of the prepared materials has been investigated using a scanning electron microscope (SEM) at 20~keV electron energy. Raman spectra of the prepared pellets were recorded with Renishaw inVia confocal Raman microscope at wavelength of 785~nm and grating of 1200~lines/mm with 1~mW laser power. Temperature dependent resistivity measurements were done using physical property measurement system (PPMS) from Quantum Design, USA.

For the electrochemical measurements, first the slurry for cathodes were prepared by mixing Na$_{0.44}$Mn$_{1-x}$Zn$_x$O$_2$ ($x=$ 0 -- 0.02) active material, PVDF (Polyvinylidene difluoride) as binder and carbon black as conductive additive in a weight ratio of 80:10:10 in NMP (N-methyl-2-pyrrolidinone) solvent followed by overnight stirring. The obtained slurry was coated on a current collector (Al foil in present case). In order to evaporate the solvent, the coated material was dried in vacuum oven at 135$^o$C for 10~hrs. After that, the circular disks of 12~mm diameter were prepared. The CR2016 coin half-cells were assembled in an nitrogen-filled glove box (Jacomex, $\leq$0.5 ppm of O$_2$ and H$_2$O level). The sodium disks prepared from sodium cubes (Sigma Aldrich, 99.9\%) with 16 mm diameter were used as counter as well as reference electrode. The glass fiber filter (Advantec, GB-100R) was used as separator and in house prepared NaClO$_4$ dissolved in ethylene carbonate (EC)/dimethyl carbonate (DMC) in a volume ratio of 1:1 was used as an electrolyte \cite{maheshMRB}. All the electrochemical characterizations were performed using VMP3 (Biologic) instrument and BTS-400 (Neware) at room temperature. The cyclic voltmmetry (CV) measurements were performed in the potential window of 1.8 -- 4.0~V vs. Na/Na$^+$ at a scan rate of 0.1/0.05~mVs$^{-1}$. The charging/discharging of the coin cell were performed in galvanostatic mode at different current densities.  

\begin{figure}[h]
\includegraphics[width=3.3in]{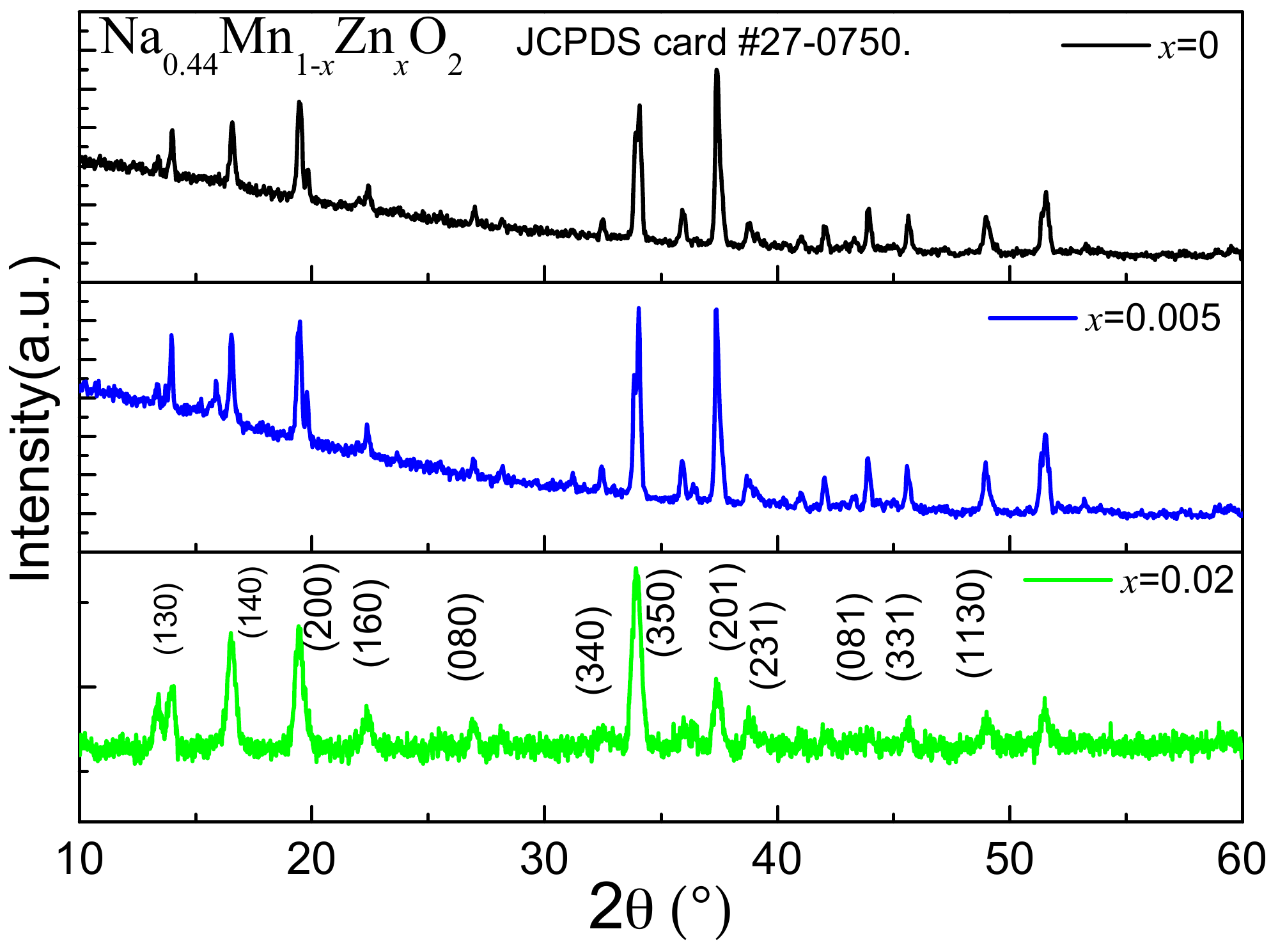}
\caption{The XRD patterns for the Na$_{0.44}$Mn$_{1-x}$Zn$_x$O$_2$ ($x=$ 0 -- 0.02) samples with peak indexing using JCPDS\# 27-0750.} 
\label{fig:Fig1_XRD}
\end{figure}

\section{\noindent ~Results and Discussion}

The x-ray diffraction (XRD) patterns for the as synthesized Na$_{0.44}$Mn$_{1-x}$Zn$_x$O$_2$ ($x=$ 0 -- 0.02) are shown in Fig.~1. The structure for $x=$ 0 sample is orthorhombic with Pbam space group (JCPDS No: 27-0750), which is in agreement with earlier reports \cite{SauvageInChem2007, Nanorods2016, NM44synRev}. The Zn substitution does not affect the structure and it remain orthorhombic for all the samples. The compositions of all the samples have been confirmed by energy dispersive x-ray spectroscopy (not shown). We have estimated the  strain and crystallite size using Williamson Hall plots \cite{WHallActa1953} from the XRD data. We found that with increasing Zn concentration, the strain increases from (2.13$\pm$0.5$)\times$10$^{-3}$ for $x$=0 to (2.6$\pm$0.2$)\times$10$^{-3}$ for $x$=0.005 and (3.2$\pm$0.3$)\times$10$^{-3}$ for $x$=0.02 smaples. This observation indicates that the substitution of Mn with the larger ion Zn might be causing this increase in internal strain. However, the crystallite size decreases from 100~nm to 50~nm with increasing Zn concentration.

\begin{figure*}
\includegraphics[width=7.0in]{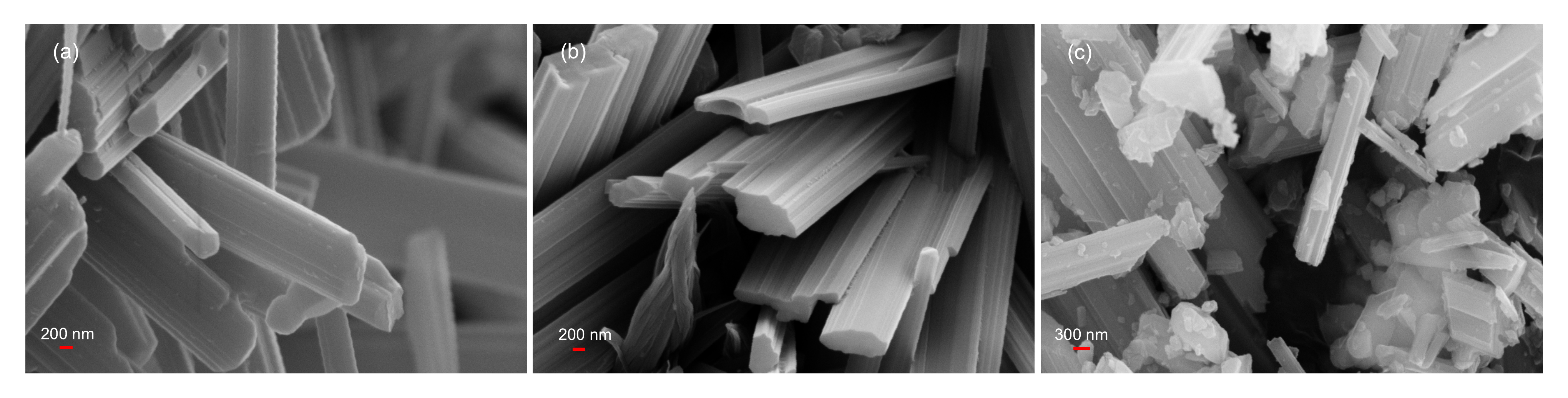}
\caption{The SEM images for Na$_{0.44}$Mn$_{1-x}$Zn$_x$O$_2$: (a) $x=$ 0, (b) $x=$ 0.005, and (c) $x=$ 0.02 samples.}
\label{fig:Fig2_SEM}
\end{figure*}

Fig.~2 shows the surface morphology of Na$_{0.44}$Mn$_{1-x}$Zn$_x$O$_2$ ($x=$ 0 -- 0.02) samples, investigated using SEM technique. In the SEM images [see Figs.~2(a-c)], we observed slabs of a few $nm$ width and $\mu$m length for Na$_{0.44}$MnO$_2$ sample. With Zn substitution, the morphology does not change much and slab structure prevails for $x=$ 0.005 and 0.02 samples. However, for $x=$ 0.02 sample there are some random shape particles also observed along with the slabs [Fig.~2(c)]. In Fig.~3, we have shown the Raman spectroscopy measurements for $x=$ 0 and 0.02 samples using a 785~nm wavelength laser. There are two major peaks observed in the Raman spectra for $x=$ 0 sample , i.e. at 654 and 369 cm$^{-1}$. These two peaks correspond to the stretching vibrations of Mn-O bonds and bending vibrations of Mn-O-Mn bonds, respectively. These peaks are slightly shifted and lower in intensity as observed at 649 and 364 cm$^{-1}$ for $x=$ 0.2 sample. Considering the diatomic approximation which assumes each metal oxygen bond as a separate and non-interacting oscillator and which has been used to explain the relation between Raman shift and bond length in many transition metal oxides \cite{WachsRamanSolidIon1991, WachsRamanJChemSoc.Faraday1996}, the shift in the Raman peaks towards lower wave number with increasing Zn concentration indicates a decrease in the force constant and thus increase in the metal oxygen bond length. Apart from these major peaks there are few minor peaks observed in both the samples. These results are consistent with earlier reports \cite{ZhaoRSC13Raman,Raman0.21}.

\begin{figure}
\includegraphics[width=3.3in]{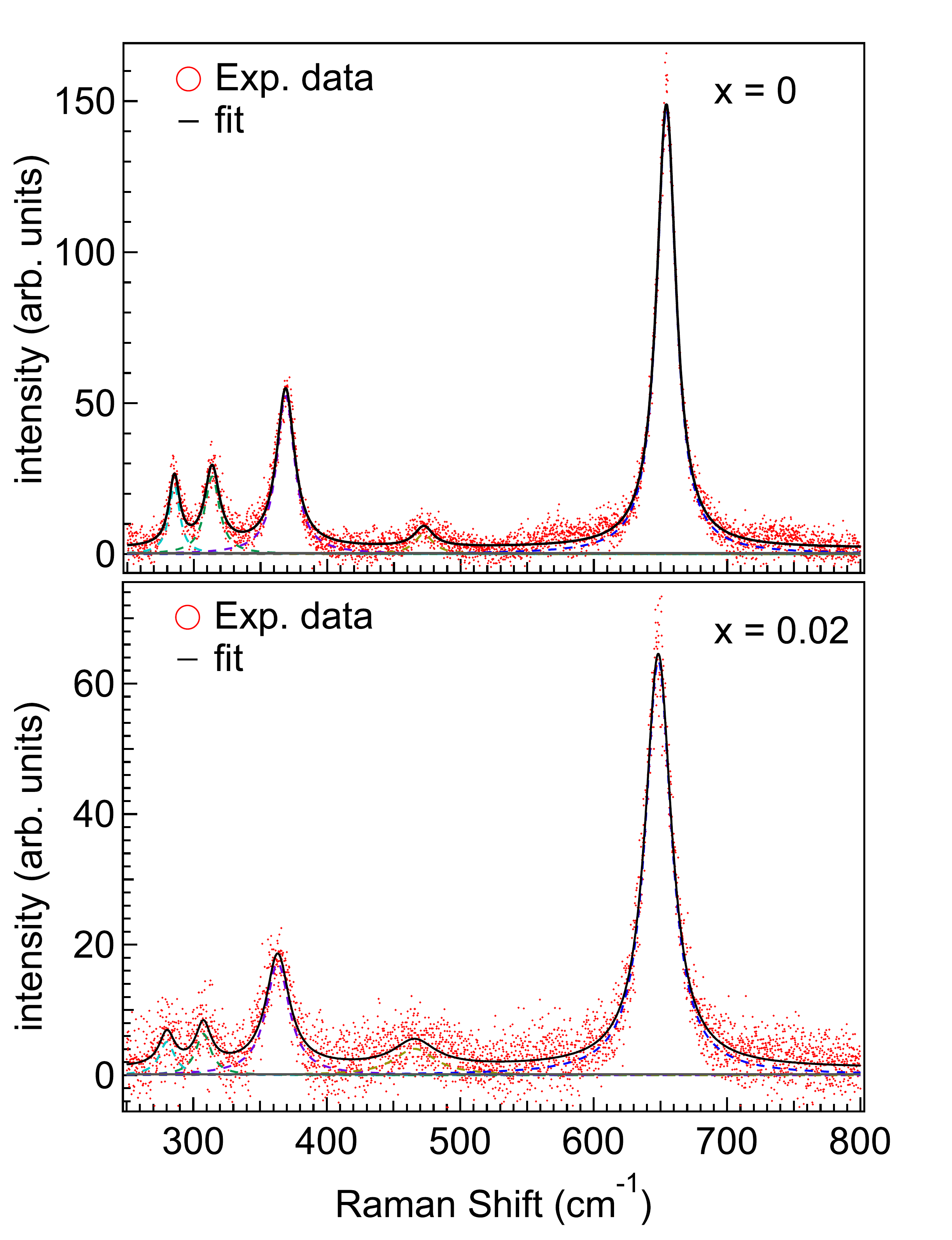}
\caption {Raman spectra of Na$_{0.44}$Mn$_{1-x}$ZnO$_2$ ($x=$ 0 and 0.02) samples, measured at room temperature using 785~nm laser. The dashed lines are fitted with Lorentzian line shape.} 
\label{fig:Raman}
\end{figure}

Now let us discuss the electronic properties which also play an important role in the electrochemical properties. The electronic band structure of 3$d$ transition metal oxides is mainly governed by the overlapping and interaction between 3$d$ orbitals of the neighboring transition metal ions. The substitution of the foreign ion alters the overall metal-metal distance, which manifest itself into the modification in the band structure. The intercalation of Na into the cathode material is correlated to the density of states near the Fermi level \cite{Molenda1 SSIon1986, XiaChemSci2018, KunduraciChemMat2006, Molenda2 RSC2017}. The high density of states (i.e. higher conductivity) results in to high rate capability, whereas the lower density of states (lower conductivity) in the cathode material leads to poor cycle life as described in the ref.~\cite{Molenda1 SSIon1986}. In a way, lattice constant, electronic properties and electrochemical properties are correlated. Therefore, in order to get an idea about how Zn substitution affects the electronic properties, we have performed the temperature dependent resistivity measurements on the three samples by four probe method. All the samples show highly insulating nature due to which we could not measure the resistivity below 280~K, as it increases beyond the measurement limit of the instrument. When compared with $x=$ 0 sample, the resistivity first decreases for $x=$ 0.005 sample and then increases for $x=$ 0.2 (Fig.~4). However, the fitting of the temperature dependent resistivity data using Arrhenius equation: $\rho=\rho_0e^{E_a/k_B T}$, where, E$_a$ is the activation energy, shows that the activation energy systematically decreases with increasing Zn substitution. For example, the activation energy is 420~meV, 360~meV and 330~meV for $x=$ 0, 0.005 and 0.2 samples, respectively. These values of the activation energies are well in agreement with the previous reports on Mn based transition metal oxides \cite{bandgap, maheshMRB}. The plausible explanation for reduction in the activation energy can be the defect levels in the band gap due to the Zn substitution. The systematic decrease in the activation energy with Zn substitution indicates narrowing the band gap and increase in density of states with doping. As mentioned above, this increase in the density of states near Fermi level in $x=$ 0.2 sample can be one of possible reason for better rate capability despite of lower specific capacity which is described in the following section.

\begin{figure}
\includegraphics[width=3.3in]{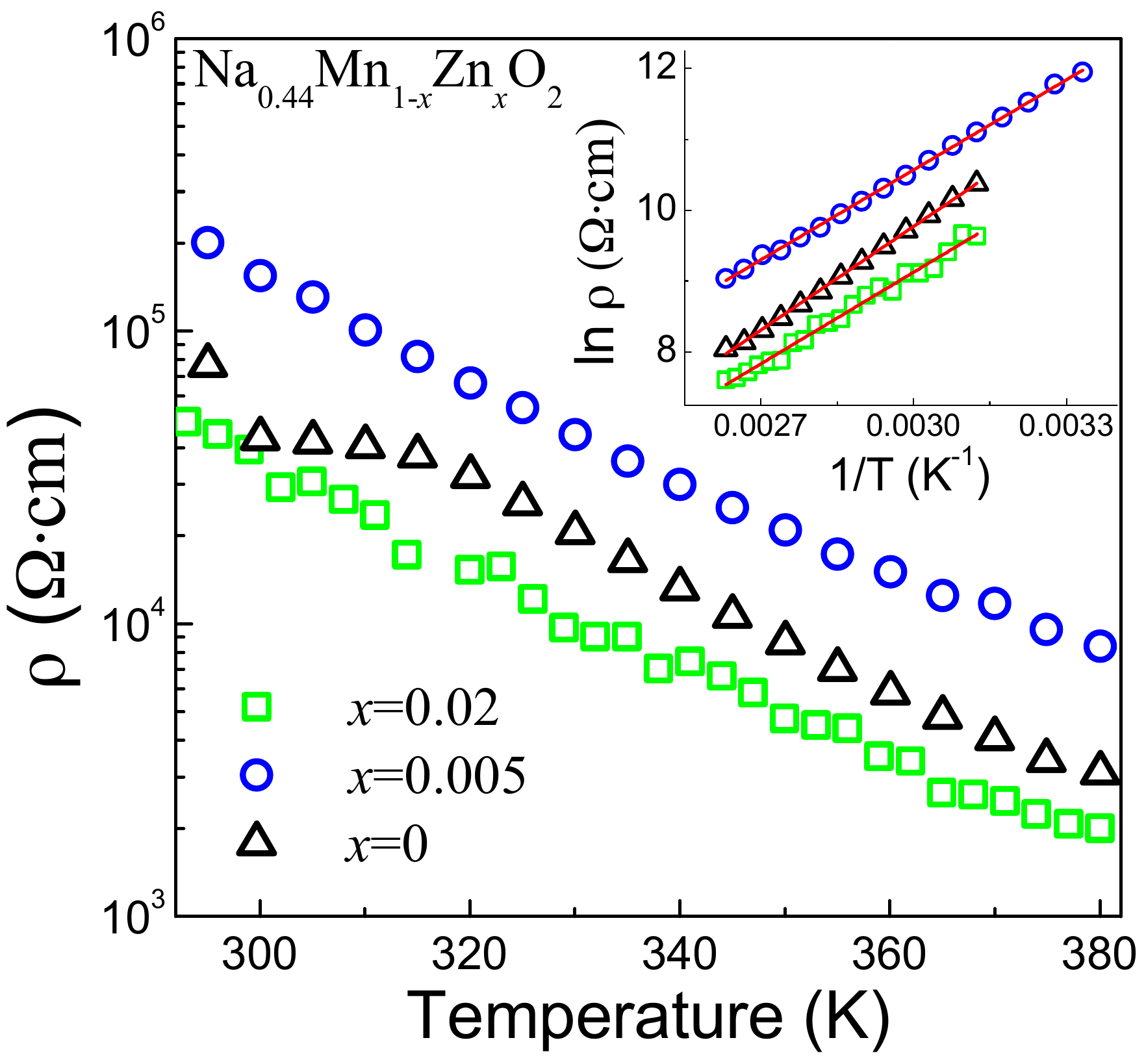}
\caption {The temperature dependent resistivity data of Na$_{0.44}$Mn$_{1-x}$ZnO$_2$ ($x=$ 0 -- 0.02). The inset shows  a plot between ln $\rho$ and 1/T with fitting by the Arrhenius equation.} 
\label{fig:Fig3_RT}
\end{figure}

The electrochemical behavior of Na$_{0.44}$Mn$_{1-x}$ZnO$_2$ ($x=$ 0 -- 0.02) as the electrode materials in Na-ion batteries has been investigated by cyclic voltametry (CV) and charging discharging measurements at different current densities on the prepared coin cells. The CV data gives information about the potentials at which the oxidation/reduction takes place, as shown in Figs.~5(a--c), the scan rate is kept as 0.1 or 0.05~mV/sec for all the CV scans. For $x=$ 0 sample, there are six distinct peaks in the oxidation, which repeats in the reduction as well as upon cycling indicating reversibility of Na ion in extraction/insertion. The multi peaks in the CV indicate a complex phase evolution where each peak corresponds to the extraction of Na from different crystallographic sites in the structure \cite{DemirelMT}. The difference between oxidation and reduction peak values is a measure of the number of free electrons (or Na-ion) involved in the charging/discharging and is given by $\Delta$E=59/n~meV (at room temperature) \cite{CV formula}, where, $\Delta$E is the difference between the oxidation and corresponding reduction peaks and n is the number of electrons involved. For n=0.44 (to extract all the Na-ion), the peak separation should be 0.13~mV. For our sample the peak separation is 0.11, 0.17, 0.17, 0.18, 0.21~mV for peaks from higher voltage to lower voltage side, which correspond to n=0.53, 0.34, 0.34, 0.32, 0.28. These values are close to the expected n = 0.44 in ideal case. The CV data for all three samples ($x=$ 0, 0.005, 0.02) are shown in Figs.~5(a--c). Now, if we look at the CV for $x=$ 0.005 and 0.02 samples, it is clear that $x=$ 0.005 sample contains all the oxidation and reduction peaks similar to that in $x=$ 0 sample with a small shift in the peak position. Moreover, the similar oxidation peaks have been observed for $x=$ 0.02 sample, but there are fewer reduction peaks when compared with the $x=$ 0 sample. In order to resolve the redox peaks for $x$=0.02 cathode we performed the CV at slower scan rate of 0.05~mV/s [Fig.~5(c)], but no significance change has been observed. The separation between the oxidation and reduction peaks has also increased indicating decrease in the number of available free electrons and irreversibility of Na ion for some sites, which affect the discharge capacity with higher Zn concentration at the Mn site. 

\begin{figure}
\includegraphics[width=3.4in]{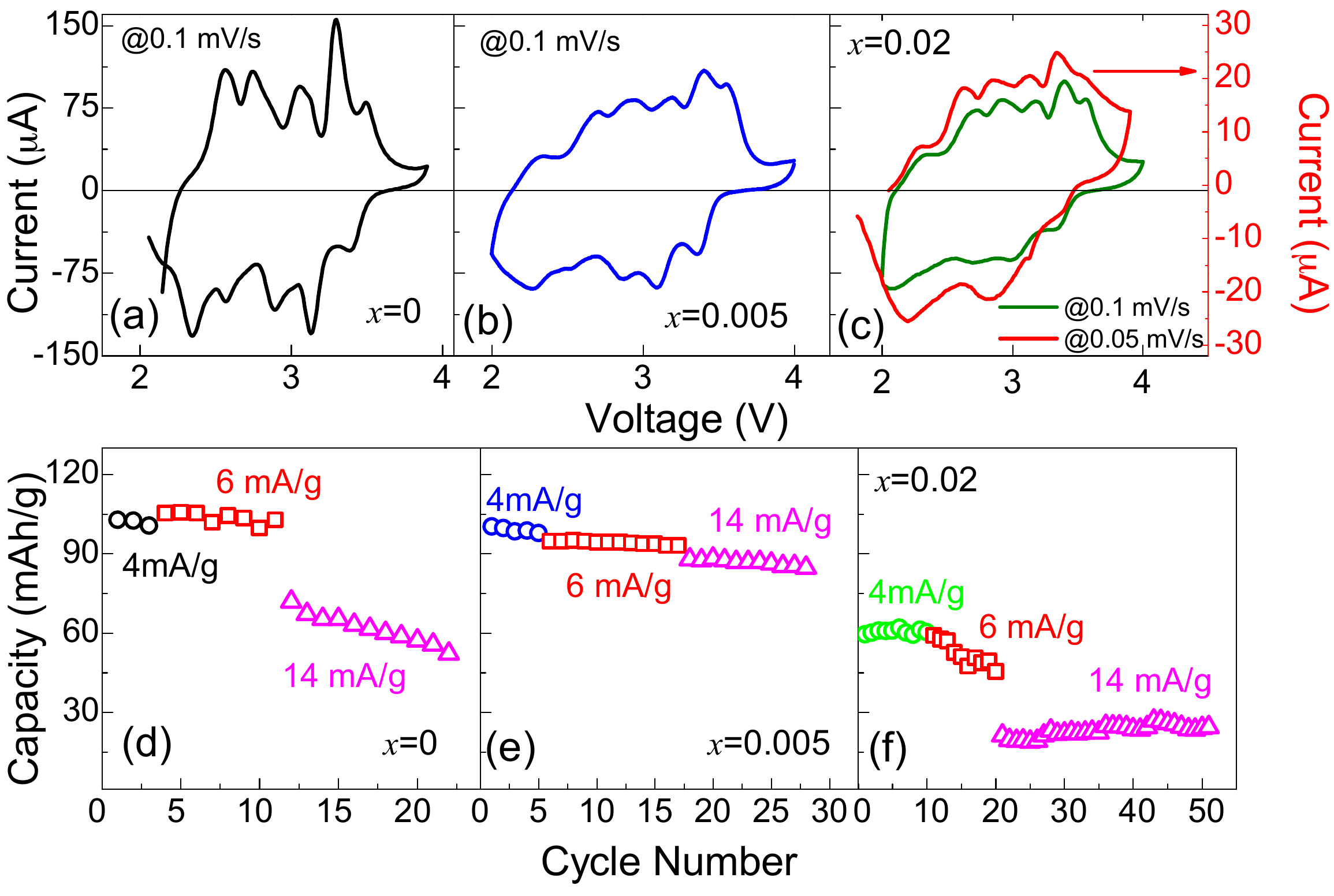}
\caption {(a--c) The cycling voltammogram and, (d--f) the capacity degradation upon cycling for Na$_{0.44}$Mn$_{1-x}$Zn$_x$O$_2$ with $x=$ 0, 0.005, 0.02, measured at different current densities.} 
\label{Fig5_new}
\end{figure}

\begin{figure}
\includegraphics[width=3.5in]{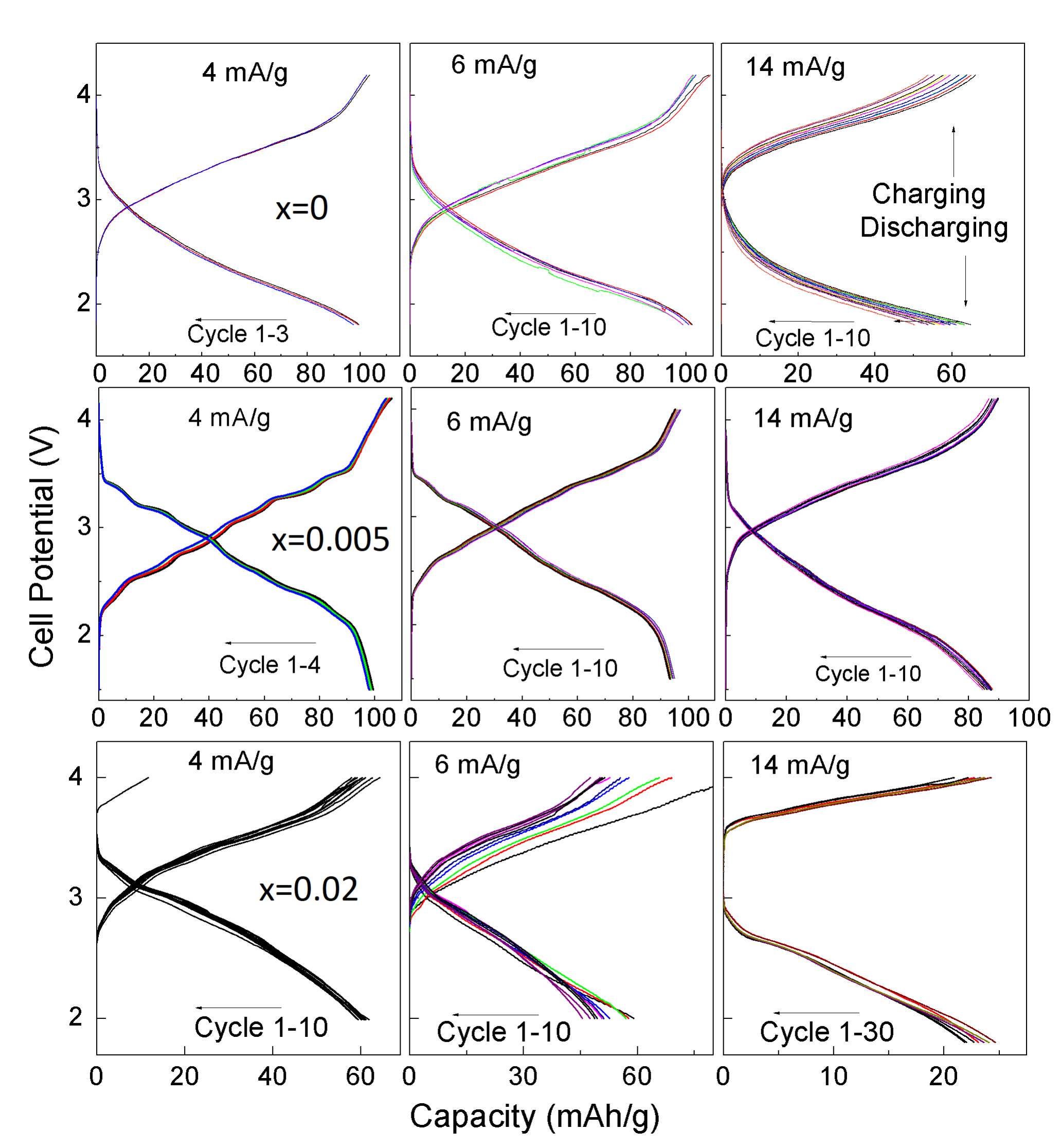}
\caption {The charging discharging cycles measured at different current densities for $x=$ 0 sample (upper panel), 0.5$\%$ (middle panel) and 2$\%$ Zn substitutions (lower panel).} 
\label{fig:Figure6_new}
\end{figure}

The galvanostatic charging-discharging curves at different current densities for Na$_{0.44}$Mn$_{1-x}$Zn$_x$O$_2$ ($x=$ 0, 0.005, 0.02) are shown in the Fig.~6 and the capacity fading with cycling is summarized in Figs.~5(d--f). For $x=$ 0 sample, the capacity is nearly 100~mAh/g (which is close to the theoretical capacity 121~mAh/g \cite{TheorycapRSC 2014}) for slow charging rates that is with current densities 4~mA/g and 6~mA/g. It can be seen that the capacity is almost constant after 10 cycles. However, for higher current density (14~mA/g) the capacity is low and degrade rapidly with cycling (nearly 25$\%$ degradation after 10 cycles), see Fig.~5(d). The possible reason for such behavior lie in the layered structure and presence of Jahn-Teller Mn$^{3+}$ ions in these samples. Similar capacity fading is also observed in other Mn based oxides where Mn has 3+ oxidation sate \cite{JTMn, maheshMRB}. As mentioned earlier the Zn substitution would decrease the concentration of Jahn-teller active Mn$^{3+}$ ions and one can expect improvement in the capacity degradation upon cycling. When comparing the capacity of $x=$ 0 and 0.005 samples after 10 cycles at 6~mA/g current density, we find that the capacity loss is nearly 4 $\%$ for the $x=$ 0 sample, whereas it is only about 1$\%$ for $x=$ 0.005 sample. In addition, the capacity is higher with improved cycling capacity at 14~mA/g current density. This clearly indicate the improvement in the cycle life with small Zn concentration (0.5\%). With further increasing Zn concentration (2\%), the overall capacity has decreases to about 60~mAh/g at slower current density of 4~mA/g. This decrease in the capacity is a manifestation of the absence of reduction peaks in the CV for 2$\%$ substituted sample, as discussed earlier. Despite of the low capacity, the durability is improved and the capacity remain unchanged after 10 cycles at 4~mA/g current density. Upon increasing the charging/discharging current density to 14~mA/g, the capacity decreases to nearly 25~mAh/g and remains constant for more than 30 measured cycles. In this way, we reveal that Zn substitution increases the cycle life possibly due to the reduction in Jahn-Teller Mn$^{3+}$ ions. Now, we discuss the possible explanation of the lower capacity in galvanostatic charging discharging for $x=$ 0.02 sample in the view of the CV measurements. The substitution of Zn$^{2+}$ can have the following effects in the parent sample: (1) it replaces the Mn$^{3+}$ (not Mn$^{4+} $, due to larger difference in ionic radii and charge) and simultaneously converts another Mn$^{3+}$ to Mn$^{4+}$ in order to maintain the charge neutrality, (2) Zn$^{2+}$ is reluctant in changing its valance state from 2+ and Mn$^{4+}$/Mn$^{5+}$ redox couple is rare and not accessible at ordinary potentials. Therefore, the presence of Zn$^{2+}$ hinders the extraction of Na from the cathode in these two ways. This hindrance clearly reflect in the CV of $x=$ 0.02 sample in terms of smaller anodic/cathodic peak current as well as less number of redox peaks, which may manifest in the low specific capacity.

\begin{figure}
\includegraphics[width=3.5in]{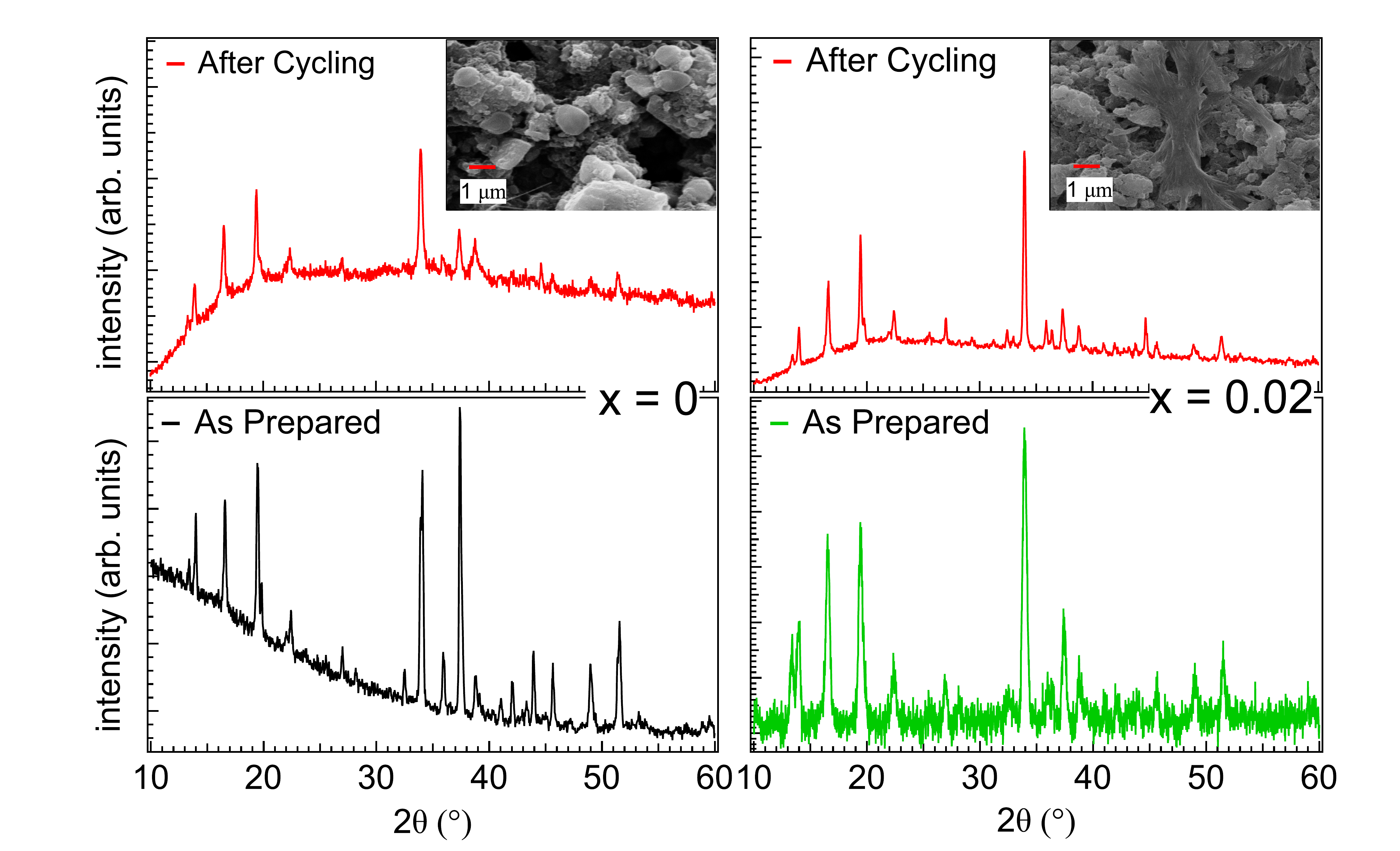}
\caption {The comparison of XRD patterns for the as prepared sample and cycled electrode of $x=$ 0 and $x=$ 0.02 samples. The insets show SEM images of the cycled electrodes.} 
\label{fig:Fig7_XRD}
\end{figure}

 In order to understand how the cycling affect the structure and morphology, we recorded the XRD patterns and SEM images of the cycled electrodes. We compare the XRD data in Fig.~7 and SEM images in the insets. It is evident from the comparison of XRD data that there are no new peaks observed after cycling and the peaks in the as prepared materials matches with those in the the cycled electrode. This shows that there is no structural or phase change taking place during/after cycling process. However, in the case of $x=$ 0 sample, the peak broadening and intensity of peaks found to different indicating a change in particle size and morphology. The SEM images of $x=$ 0 and  $x=$ 0.02 electrodes show that the electrode surface for $x=$ 0 is porous with large voids as compared to that for $x=$ 0.02 electrode, which might be responsible for rapid capacity degradation upon cycling.

\section{\noindent ~Conclusions}

In conclusion, we have synthesized nano-size rod shaped Na$_{0.44}$Mn$_{1-x}$Zn$_x$O$_2$ ($x=$ 0 -- 0.02) via sol gel method and studied the physical properties and electrochemical performance for Na-ion batteries. The structure is orthorhombic (Pbam space group) and it does not change with Zn substitution. All the samples are highly insulating and the activation energy, deduced from Arrhenius fitting, decreases with higher Zn concentration. The specific capacity for Na$_{0.44}$MnO$_2$ is nearly 100~mAh/g, which is close to the theoretically predicted value. The capacity does not increase with Zn substitution, but we observed significant improvement in cycling life. The capacity reduction is only 1$\%$ for the 0.5\% Zn sample after 10 cycles as compared to the 4\% of $x=$ 0 sample, measured at 6~mA/g current density. Our study reveal that Zn substitution changes Mn ion from 3+ to 4+ valence state and improve the battery life. We observed changes in the morphology of electrodes after cycling.

\section*{\noindent ~Acknowledgments}

RSD acknowledges the financial support from SERB-DST [Early Career Research (ECR) Award, project reference no. ECR/2015/000159], BRNS (DAE Young Scientist Research Award, project sanction no. 34/20/12/2015/BRNS), and IIT Delhi (FIRP project no. MI01418). Mahesh and Rishabh thank SERB-DST (NPDF, no PDF/2016/003565), and DST-inspire, respectively, for the fellowship. Rakesh thanks IIT Delhi for postdoctoral fellowship through FIRP project. The authors acknowledge central research facility (CRF) and the physics department of IIT Delhi for providing research facilities: XRD, SEM, Raman, and PPMS EVERCOOL--II.

\end{document}